# Electrostatic co-assembly of magnetic nanoparticles and fluorescent nanospheres: a versatile approach toward bimodal nanorods

Jérôme Fresnais[a], Eléna Ishow[b], Olivier Sandre[a] and Jean-François Berret[c,@]

*(a)* : Laboratoire de Physico-chimie des Electrolytes, Colloïdes et Sciences Analytiques, UMR 7195, UPMC Univ Paris 6 / CNRS / ESPCI 4 place Jussieu, 75252 Paris Cedex 05, France. *(b)* : Laboratoire de Photophysique et Photochimie Supramoléculaires et Macromoléculaires, UMR CNRS 8531, ENS Cachan, 61 Avenue du Président Wilson, 94235 Cachan, France. *(c)* : Matière et Systèmes Complexes, UMR 7057 CNRS Université Denis Diderot Paris-VII Bâtiment Condorcet 10 rue Alice Domon et Léonie Duquet, 75205 Paris France

The elaboration of multimodal nanoparticles stimulates tremendous interest owing to their numerous potentialities in many applicative fields like optoelectronics,[1] photonics[2] and especially bioimaging.[3-5] The concomitant association of various properties (optical, electrochemical, magnetic) allows for the use of complementary stimuli in order to probe the interactions between the nanoparticles and their surroundings.[6,7] Nanoparticles (NPs) have thus become highly praised tools to image cells and tissues with a large contrast compatible with the dimensions of biological materials and the existence of quantum confinement effects induced by the reduced dimensions. In this context, the combination of magnetism and emissive properties such as fluorescence appears particularly attractive for non-invasive investigations[8-10], cell sorting or drug vectorization.[11] Therefore, combining both fluorescence and magnetism requires the delicate construction of hybrid assemblies.[12-14] Most of the magnetic nanoparticles are made of metallic oxides or alloys, *e.g.* $\gamma$-Fe$_2$O$_3$, Fe$_3$O$_4$, FePt, while the target fluorescent entities are often organic dyes or quantum dots (QDs).[15]

The most common route to combine both partners in a single system was based on their covalent association through targeted surface functionalization. However this required tedious and long synthetic steps, especially for QDs to circumvent issues of cytotoxicity, photooxidation and instability in aqueous media. A second route involves the incorporation of magnetic and emissive nanoparticles into the same polymeric or silica scaffolds,[15-16] which results in an enhanced photostability of the organic or inorganic dyes.[17-19] Yet the physical dilution of the active units led to a decrease in the material response efficiency and loss of possible cooperativity. A third alternative approach would consist in establishing cohesive electrostatic interactions between the components through a binding agent. This approach appears particularly attractive for two reasons: first its versatility and second the high density of active components present in the final material. In this context, we have recently demonstrated that 7 nm maghemite ($\gamma$-Fe$_2$O$_3$) nanoparticles could successfully be assembled through electrostatic interactions promoted by a cationic copolymer and $\gamma$-Fe$_2$O$_3$ nanoparticles coated by poly(acrylate) units. Indeed, when dialysis of the reaction mixture down to low ionic strength (10$^{-3}$ mol L$^{-1}$) was performed under a static magnetic field, the components got reproducibly associated into 1 to 50 $\mu$m long rods where the superparamagnetic properties of the elementary 7 nm units remained unchanged.[20, 21] We thus decided to take advantage of this assembling process to incorporate fluorescent dyes during the association step. This apparently simple strategy needs however to address numerous issues. The dyes must emit in the red, far from the strong absorption region of the magnetic nanoparticles to avoid reabsorption effects. They must present a large density of fluorophores to produce highly light-sensitive systems. Finally they must possess negative charges on their surface in order to interact with the "gluing" cationic copolymer. It turns out that organic nanoparticles made of non-doped small molecules has recently emerged as very promising materials, especially in the field of one and two-photon emission where a high load in active dyes enables sensitive detection[22] and structural confinement has been observed.[28-30]

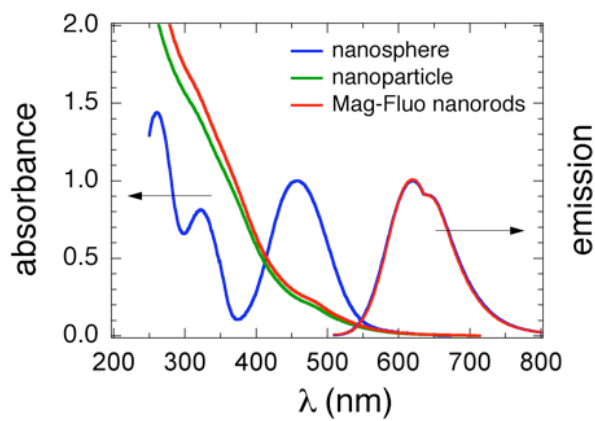

**Figure 1** : *Left scale* : normalized absorbance spectra of nanospheres (blue), nanoparticles (green), and magnetic fluorescent rods (red). *Right scale* : normalized emission spectra of nanospheres (blue) and magnetic fluorescent rods (red). Note that both nanospheres and magnetic fluorescent rods emission spectra are superimposed.

We report herein that the non-covalent association of fluorescent organic nanospheres and magnetic nanoparticles can represent a novel and elegant approach to elaborate bimodal nano-objects that exhibit both superparamagnetic and emissive properties. Since such supramolecular assemblies occur in the very final step, this route offers high versatility and considerable spare of synthetic work, not to mention its large applicability to other functional systems.

The formation of magneto-fluorescent nanorods proceeded first through the fabrication of anionic magnetic and fluorescent nanoparticles whose size, density of charge and concentration were controlled. The bimodal co-assembly driven by electrostatic interactions was then realized by mixing both the magnetic and fluorescent colloids in the presence of a third component, a cationic polyelectrolyte to serve as a "glue" for the whole.

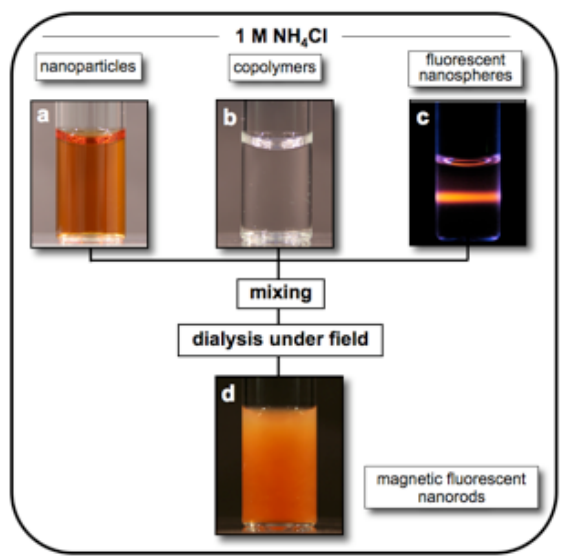

**Figure 2** : Schematic scheme developed for the co-assembly of nanoparticle, nanosphere and polymer in aqueous solution. VIAL **a** : PAA$_{2K}$-coated 7 nm γ-Fe$_2$O$_3$ nanoparticles; VIAL **b** : PTEA$_{11k}$-b-PAM$_{30k}$; VIAL **c** : PAA$_{2K}$-coated fluorescent nanospheres made from FVIN. All dispersions are at 1 mol L$^{-1}$ NH$_4$Cl; VIAL **d** : dispersion of fluorescent superparamagnetic nanorods.

We retained superparamagnetic maghemite nanoparticles (γ-Fe$_2$O$_3$) which have thoroughly been studied during the last decade with respect to some fundamental properties: *i)* the control of the size distribution[23] and *ii)* the colloidal stability and the modification of the surface chemistry of the particles.[24] The ferrofluid dispersion, obtained from the alkaline coprecipitation of iron(II) and iron(III) salts after the protocol of Massart et al.,[23] was characterized by a log-normal size distribution, with a median diameter 7.1 nm and a polydispersity 0.26.[25] As synthesized, the particles were positively charged, with nitrate counterions adsorbed on their surfaces. The maghemite NPs were subsequently coated by poly(acrylic acid) chains of molecular weight of 2000 g mol$^{-1}$ (hereafter abbreviated as PAA$_{2K}$), using the precipitation-redispersion method.[26] The thickness of the PAA$_{2K}$ adlayer (2 - 3 nm) was determined by dynamic light scattering by comparing the bare and coated particles (see Supporting Information, Figure SI-2a). This complexation method allowed the formation of anionically charged NPs, stabilized by electrosteric interactions between pH 6 and 11 and for ionic strength up to 1 mol L$^{-1}$ (NH$_4$Cl) (Figure SI-2b).[27]

As for the fluorescent systems, we turned our attention to push-pull triarylamine derivatives that we recently synthesized and could precipitate into nanoparticles, strongly emitting from blue to red in water (see Supporting Information, Figure SI-1). Since the emission reabsorption by the γ-Fe$_2$O$_3$ particles, absorbing in the visible, must be circumvented, we selected the red-emitting derivative 4-di(4'-tert-butylbiphenyl-4-yl)amino-4'-dicyanovinylbenzene FVIN whose fluorescence spectrum peaked at 632 nm in the solid state far from the γ-Fe$_2$O$_3$ absorption range .[22]

The fabrication of the corresponding fluorescent particles was done by the rapid mixing of a small amount of FVIN stock solution (c = 0.05 wt. %) dissolved in acetone with a large excess of water. For the nanospheres to be assembled by the "gluing" cationic polymers, a 1 wt. % solution of PAA$_{2K}$ (instead of deionized water) was used in the reprecipitation process. The same polymer was employed as that for the coating of the maghemite nanoparticles for obvious compatibility reasons. Remarkably spherical and monodisperse colloids, dubbed nanospheres in the following were obtained, as evidenced by transmission electron and optical microscopy (see Supporting Information, Figure SI-3). Determination of hydrodynamic diameters gave $D_H$ = 180 nm. As expected, uncoated particles exhibited slightly reduced $D_H$ equal to 150 nm. These results agreed remarkably well with the size measurements obtained by transmission electron microscopy (TEM). The electrophoretic mobilities ($\mu_E$) of the PAA$_{2K}$-coated nanospheres were found to be $\mu_E$ = -5.7×10$^{-4}$ cm$^2$ V$^{-1}$ s$^{-1}$, which shows that the system was strongly anionically charged. Compared to the value for uncoated nanospheres ($\mu_E$ = -3.2×10$^{-4}$ cm$^2$ V$^{-1}$ s$^{-1}$), the increase in $\mu_E$ was assigned to the adsorption of the $PAA_{2K}$ at the nanosphere-solvent interface (Figure SI-2b). The colloidal stability of the coated nanospheres in aqueous solution could be proven for a period longer than 2 months at room temperature.

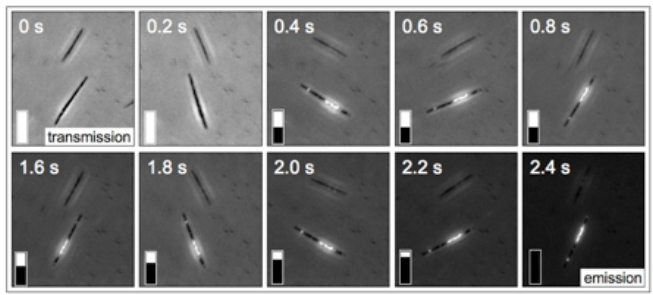

**Figure 3** : Transmission and emission images of magnetic and fluorescent rods in water, sealed between glass plates. The vertical bar corresponds to 10 µm and indicates the proportions between the white light and the 465 nm-illumination intensities.

The UV-visible absorption spectra of $PAA_{2K}$-coated nanospheres exhibited two bands in the blue region and an additional band in the visible region centered at 468 nm (Figure 1). Based on previous experimental and computational studies (time-dependent density functional theory), this latter band has been ascribed to a charge transfer transition from the amino group to the dicyanovinylene unit forming a radiative excited state. Excitation in the UV or in the visible led to the same broad emission centered at 632 nm, revealing a large Stokes shift of 164 nm. These values were insensitive to the physico-chemical conditions of the dispersions (temperature, pH) and identical to those obtained for uncoated nanospheres. The emission spectrum of the aqueous dispersions of $PAA_{2K}$-coated nanospheres presented a weak spectral overlap with the absorption band of the $\gamma$-$Fe_2O_3$ dispersions, validating the choice made of both partners on a spectral basis.

The combination of the fluorescent and magnetic particles was performed after the same protocol as that previously used for getting giant magnetic assemblies.[20] The adopted strategy involved in a first step the preparation of three separate 1 mol $L^{-1}$ $NH_4Cl$ solutions containing respectively of *i)* 7 nm-sized $\gamma$-$Fe_2O_3$ nanoparticles, *ii)* 180 nm fluorescent nanospheres and *iii)* a cationic polymer. The polymer consisted in poly(trimethylammonium ethylacrylate methyl sulfate-*b*-poly(acrylamide) block copolymer (abbreviated $PTEA_{11K}$-*b*-$PAM_{30K}$) with molecular weights 11000 and 30000 g $mol^{-1}$ for the charged and neutral segments, respectively.[31] The concentrations were adjusted to 0.2 wt. % for the magnetic NPs and cationic $PTEA_{11K}$-*b*-$PAM_{30K}$ and 0.01 wt. % for the FVIN nanospheres. In a second step, mixing of the three solutions in a ratio 1/0.2/5 ($PAA_{2K}$–$\gamma$-$Fe_2O_3$/FVIN/$PTEA_{11K}$-*b*-$PAM_{30K}$) was done under stirring. Dynamic light scattering measurements were performed at each step of the formulation and revealed no association between the different components. In a third step, the ionic strength of the mixture was progressively diminished by dialysis against deionized water (figure 2) in the presence of an external magnetic field (0.3 T) until a stationary state was reached ($I_S = 10^{-3}$ mol.$L^{-1}$). In this way, as already demonstrated,[20] the electrostatically screened polymer and nanoparticle system underwent an abrupt transition between an unassociated and a cluster state, producing supracolloidal assemblies. At this stage, it is assumed that this phenomenon was driven by the desorption-adsorption of the polymers onto the particle surfaces.

A drop of the dialyzed mixture was sealed between two glass plates and imaged by optical microscopy (Figure 3). Transmission imaging displayed elongated aggregates, with length distribution described by a log-normal function with a median length of 10 µm and a polydispersity of 0.40 (Figure SI-4). The diameter of the rods however could not be estimated by optical microscopy.

A movie of two rods subjected to a rotating magnetic field (0.63 Hz) has been recorded and it is shown in Supporting Information. In Figure 4, two snapshots from times 0 – 0.8 s and 1.6 - 2.4 s corresponding each to a 180° turn of the field depict the rotation of the rods, indicating that the rods are magnetically active. By analogy with previous work, we assume that these rods have inherited the superparamagnetic properties from their constituting maghemite nanoparticles. During the rotation, the excitation at 465 nm was switched on at 0.4 s. Its intensity was progressively increased, while the white light used for transmission imaging was dimmed. Bright areas within the bottom rotating rod appeared under illumination at 465 nm, proving the incorporation of the fluorescent nanospheres within the magnetic structures.

Steady-state fluorescent experiments were performed on water suspensions of magneto-fluorescent nanorods. Excitation at 465 nm, at the red edge of the maghemite absorption band and close to the FVIN absorption maximum produced an emission spectrum which superimposed that of FVIN nanospheres dispersed in deionized water or $PAA_{2K}$-coated. This overtly evidenced a lack of significant electronic interactions between the organic material and its inorganic surrounding, and proved our concept of non-covalent self-assembling to get multifunctional architectures with orthogonal properties.

To gain insight into the microscopic structure of the rods, TEM imaging was performed. One clearly observed cylindrical structures with a high aspect ratio, where the 7 nm-nanoparticles form a tight and homogeneous ensemble. The rod displayed in Figure 4a has a diameter of 200 nm for a length of 2.25 mm. The volume fraction occupied by the nanoparticles was assessed by small-angle neutron scattering to be 25 % of the whole.[20] This corresponds to $5 \times 10^4$ nanoparticles per micrometer for rods with a diameter of 200 nm.

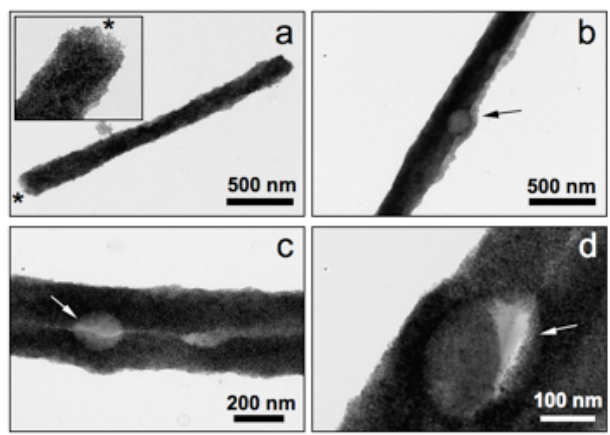

**Figure 4**: TEM images of a magnetic rod (a) and magnetic fluorescent rods (b, c, and d) with different magnifications (×3000 for a and b, ×16000 for c and ×32000 for d) The arrows indicate the fluorescent nanospheres embedded in the rod.

TEM imaging of magneto-fluorescent rods evidenced the same elongated structures where this time *circa* 200 nm sized-globules were embedded. This dimension matched the hydrodynamic diameter of the fluorescent FVIN nanospheres found at $D_H = 180$ nm. In figures 4b-d, the globules and the metallic nanoparticles forming the rods exhibit a marked difference in the electronic contrast, indicating that the former should be made of organic matter. This reinforces our assumption of FVIN nanospheres incorporated into maghemite nanorods. Figures 4c and 4d show moreover a partial deformation of the nanospheres along the axial direction of the rod. This elongation might be due to strong mechanical constraints arising during the assembling process, as well as to the relative softness of the nanospheres. As already observed by fluorescence microscopy, the nanospheres were randomly distributed in the bulk of the rods, with a preferential localization in the central (instead of apical) part of the rods. These data show that the magnetic field driven assembly of the $\gamma$-Fe$_2$O$_3$ nanoparticles were not impeded by the presence ~ 20 times larger organic colloids. The understanding of the growth process in the presence of large entities is under current progress.

In conclusion, highly anisotropic rods containing anionic magnetic nanoparticles and organic fluorescent nanospheres were built using an electrostatic co-assembling strategy in the presence of cationic polyelectrolytes. TEM and optical microscopy showed typical lengths between 1 and 50 mm, and mean diameters around 200 nm. The hybrid elongated architectures maintain the superparamagnetic properties of the $\gamma$-Fe$_2$O$_3$ constituents, as evidenced by their in-phase response to an external rotating magnetic field. Fluorescence could be produced by the embedded nanospheres which kept emitting in the red region far from the strong absorption band of iron oxide. The approach reported here offers an easy and versatile way of elaborating multimodal nanomaterials. In terms of applications, these magneto-fluorescent rods can serve as micrometric actuators and sensors for microfluidics and microrheology, as well as addressable agents for bioimaging.

### Experimental Section

Experimental details on the nanoparticle synthesis and the optical, structural and magnetic characterizations of the nanorods and their constituents are given in the supporting information. .[21, 24, 25, 33-35]

### Acknowledgments

We are very grateful to Jacques Servais from MSC-Université Paris Diderot for the design and realization of the home-made rotating field device and electronics; to Benoit Ladoux (MSC) for access to microscopy and imaging facility; to Aude Michel (PECSA-UPMC-Université Paris 6) for the TEM and micro-diffraction experiments. This research is supported by the Agence Nationale de la Recherche, under the contract BLAN07-3_206866 and by the European Community through the project: "NANO3T—Biofunctionalized Metal and Magnetic Nanoparticles for Targeted Tumor Therapy", project number 214137 (FP7-NMP-2007-SMALL-1). This research was supported in part by Rhodia.

**TOC figure**

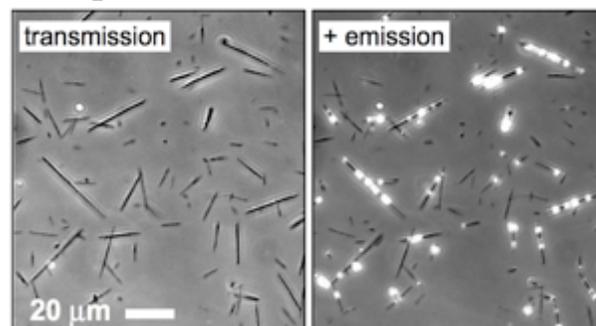

*Electrostatic co-assembly of magnetic nanoparticles and fluorescent nanospheres: a versatile approach toward bimodal nanorods*

## Supporting Information

Electrostatic co-assembly of magnetic nanoparticles and fluorescent nanospheres: a versatile approach toward bimodal nanorods

*Jérôme Fresnais, Eléna Ishow, Olivier Sandre and Jean-François Berret*

*Outline*

**SI-1 – Formula and emission spectra of fluorescent molecules**
**SI-2 – Characterization of particles and nanospheres**
**SI-2.1 –** Dynamic light scattering and electrophoresis
**SI-2.2 –** Optical microscopy and TEM
**SI-3 – Characterization of the magneto-fluorescent nanorods**
**SI-4 – Movies**
**Movie#1** : Brownian nanorods as seen by transmission and emission optical microscopy. No magnetic field was applied.
**Movie#2** : Magneto-fluorescent nanorods submitted to a rotating mangnetic field oriented by a magnetic of amplitude 10 mT and frequency 0.63 Hz.

**SI-1 – Formula and emission spectra of fluorescent molecules**
The molecule formula is given in the figure SI-1. Note that the emission wavelength depends strongly on the final end group bounded on the molecule. We selected the 4-di(4'-tert-butylbiphenyl-4-yl)amino-4'-dicyanovinyl benzene FVIN molecule for its emission wavelength at 632 nm.

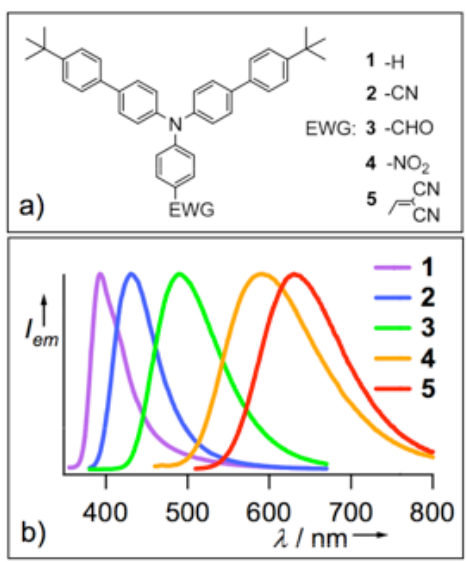

**Figure SI-1 :** a) Structure of push-pull triarylamine molecules emitting in the solid state from blue (compound 1) to red (compound 5) when changing the strength of the electron-withdrawing group. b) Normalized emission spectra ($\lambda_{exc}$ = 343 nm) of 200 nm nanospheres made from compounds 1 - 5.

**SI-2 – Characterization of particles and nanospheres**

*SI-2.1 – Dynamic light scattering and electrophoresis*

The hydrodynanic diameter distributions for the colloids and polymers were obtained using a Zetasizer Nano ZS (Malvern Instrument) operating at the scattering angle 174°. The collective diffusion coefficient D was determined from the second-order autocorrelation function of the scattered light. From the value of the coefficient, the hydrodynamic diameter of the colloids was calculated according to the Stokes-Einstein relation, $D_H = k_B T/(3\pi\eta_S D)$, where $k_B$ is the Boltzmann constant, T the temperature (T = 298 K) and $\eta_S$ the solvent viscosity ($\eta_S$ = 0.89×10$^{-3}$ Pa s for water). FigureSI-1a shows the intensity distributions as function of $D_H$ for the PAA$_{2K}$-coated magnetic particles and fluorescent nanospheres. The data for the uncoated nanospheres were included for comparison. The average $D_H$ values for the three dispersions are listed in Table SI-I. Note that the $D_H$ for PAA$_{2K}$–γ-Fe$_2$O$_3$ was 5 nm above that the uncoated particles (*i.e.* 14 nm).[25] It should be mentioned here that the hydrodynamic sizes appear larger than those determined by TEM. The reasons for that are well-known : *i)* the particles are slightly anisotropic (aspect ratio 1.2); *ii)* when particles are distributed, as this is the case here, light scattering is sensitive to the largest particles of the distribution.

Measurements of the electrophoretic mobility $\mu_E$ were carried out on the same three dispersions on the same instrument. Using laser Doppler velocimetry, the technique is based on the Phase Analysis Light Scattering (PALS) method. The electrophoretic light scattering intensities have been plotted as a function of the mobility in Figure SI-1b, whereas the m$_E$-values at the peak maximum are given in the main manuscript and in Table SI-I.

| colloids | $D_H$ nm | z mV | $m_E$ cm$^2$V$^{-1}$s$^{-1}$ |
|---|---|---|---|
| PAA$_{2K}$-γ-Fe$_2$O$_3$ | 19 | -48 | -3.8×10$^{-4}$ |
| PAA$_{2K}$-FVIN | 180 | -70 | -5.7×10$^{-4}$ |
| FVIN | 150 | -41 | -3.2×10$^{-4}$ |

**Table SI-1** : Hydrodynamic diameter ($D_H$), zeta potential (z) et electrophoretic mobility ($\mu_E$) for coated particles and nanospheres investigated in this work.

*SI-2.2 – Optical microscopy and TEM*
Phase-contrast and fluorescence images were acquired on an IX71 inverted microscope (Olympus) equipped with a 60X objective. We used a Photometrics Cascade camera (Roper Scientific) and Metaview software (Universal Imaging Inc.) for the image acquisition. Further treatment was achieved using the ImageJ software (http://rsb.info.nih.gov/ij/).

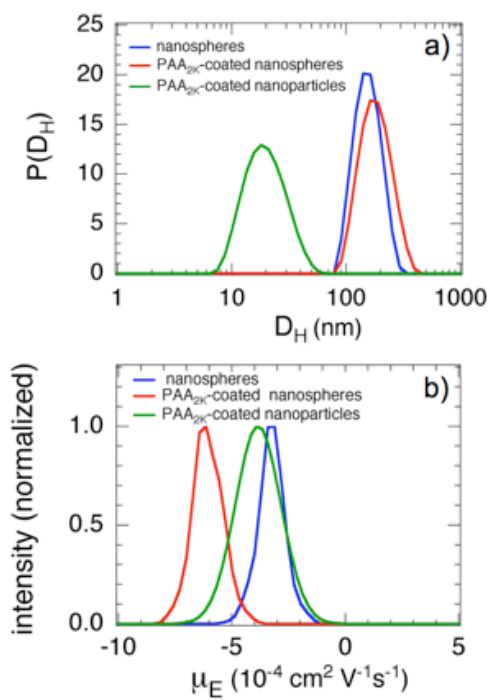

**Figure SI-2** : a) Distribution of hydrodynamic diameters $D_H$ obtained by dynamic light scattering for dispersions of $PAA_{2K}$-coated nanoparticles and nanospheres. The data for uncoated nanospheres are shown for comparison. b) Electrophoretic data for the same three colloids.

An aqueous dispersion of nanospheres was investigated in emission at the wavelength 465 nm, that is close to the absorption maximum of the FVIN fluorophore. The nanospheres in Figure SI-3 appeared as yellow-orange dots in a red dark background. The field of view of Figure SI-3 covers the range 45×50 mm$^2$.

An accurate description of the nanosphere microstructure was published recently.[22] We recall here in the inset a TEM image of the fluorescent colloids obtained by the precipitation technique. The diameter of the sphere is 120 nm. For the TEM experiments, we acknowledge the technical assistance by Aude Michel from the Laboratoire de Physico-chimie des Electrolytes, Colloïdes et Sciences Analytiques for an electron beam microdiffraction experiment with a Jeol-100 CX transmission electron microscope at the SIARE facility of University Pierre et Marie Curie (Paris 6).

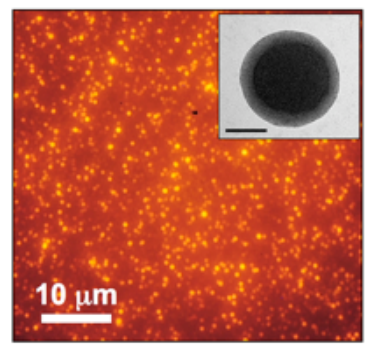

**Figure SI-3** : Emission image of a nanosphere dispersion obtained by optical microcopy ($l_{exc}$ = 514 nm). Inset : TEM image of one fluorescent colloid (bar : 50 nm).

**SI-3 – Characterization of the magneto-fluorescent nanorods**

Figure SI-4 displays the length distribution of the magneto-fluorescent nanorods. The data were those of Figure 4. The statistics was acquired on 123 specimens, yielding a distribution centered around 10 µm. The data were fitted by a log-normal distribution function with median length $L_{Rod}$ = 9.97 ± 0.5 µm, and a polydispersity s = 0.40. These data are reported in the main text.

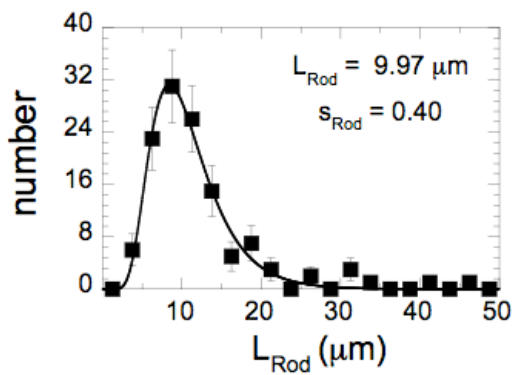

**Figure SI-4** : *Length distribution of nanorods observed in Figure 4. The continuous line was derived from best fit calculation using a log-normal distribution.*